\documentclass[preprint,12pt]{elsarticle}

\usepackage{amssymb}
\usepackage{amsmath}
\usepackage{float}
\journal{Optics $\&$ Laser Technology}
\DeclareUnicodeCharacter{202F}{~}
\begin{document}

\begin{frontmatter}



\title{High Dynamic Range enhancement in Mueller matrix polarimetry} 


\author[a,b]{Lourdes Camblor-Navarro} 
\author[a,b]{Iago Pardo}
\author[a,b]{Oriol Arteaga}

\affiliation[a]{organization={Dep. Física Aplicada, PLAT group, Universitat de Barcelona},
            city={Barcelona},
            postcode={08028}, 
            country={Spain}}
\affiliation[b]{organization={Institute of Nanoscience and Nanotechnology (IN2UB), Universitat de Barcelona},
            city={Barcelona},
            postcode={08028}, 
            country={Spain}}

\begin{abstract}
Mueller matrix (MM) polarimetry is an effective, non-invasive tool for retrieving information from complex media. However, the finite dynamic range of optical detectors poses a significant challenge when measurements involve strong intensity contrasts, where bright regions risk saturation while dark regions suffer from poor signal-to-noise ratio. To address this challenge, this article presents a straightforward, high dynamic range methodology that does not require non-linear algorithms. The proposed technique relies on the direct addition of raw intensities captured at multiple exposure times prior to the calculation of the MM. By extending the effective well-depth of the detector, this technique allows the 16 MM elements to be calculated across different hardware configurations with a significantly improved signal-to-noise ratio in low-intensity regions while eliminating artifacts caused by saturation. This approach offers a simple yet efficient solution for the characterization of  samples, eliminating the need for hardware modifications or software trade-offs.
\end{abstract}

\begin{figure}[H]
    \centering
    \includegraphics[width=5in, height=2in]{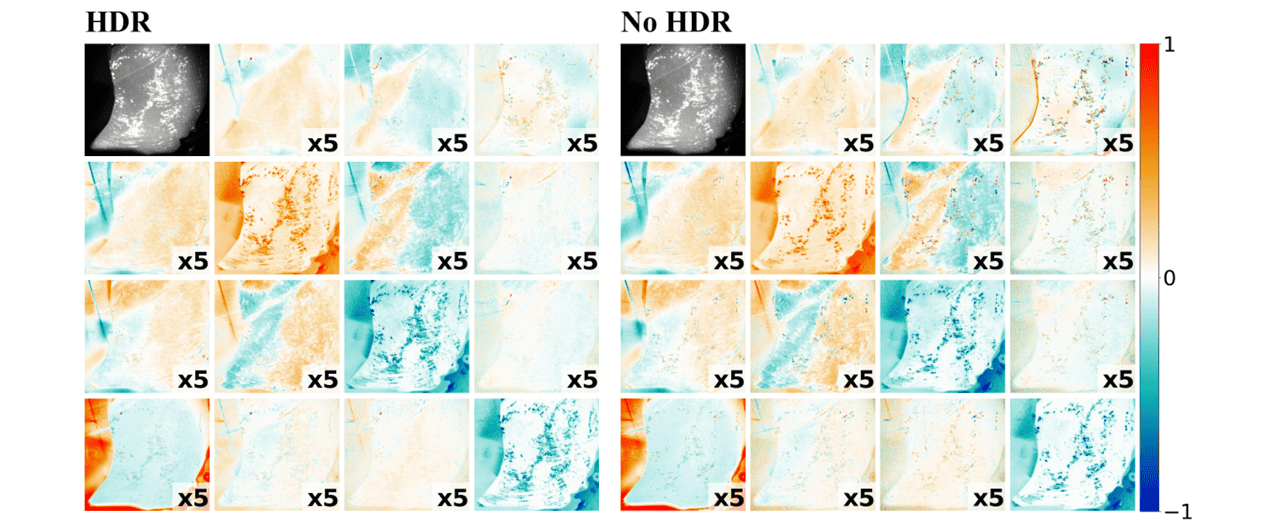}
\end{figure}

\begin{highlights}
\item Overcomes the limitation of sensor dynamic range.
\item Improves Signal to Noise Ratio and reduces saturation artifacts.
\item Does not need non-linear algorithms.
\item Tested in both Mueller matrix spectroscopic ellipsometry and Mueller matrix imaging.

\end{highlights}


\end{frontmatter}



\section{Introduction}
\label{sect:intro}  

Over the past decades, Mueller matrix (MM) polarimetry has been proven to be an effective, non-invasive tool for retrieving information from complex media \cite{he2018mueller, ghosh2008mueller}. As it provides the complete polarimetric response of a sample (comprising all 16 MM elements), techniques involving spectroscopic and imaging measurements such as MM ellipsometry and MM imaging, respectively, have gained significant relevance across different fields, ranging from biomedicine \cite{wang2016integrated, khan2023characterization} to material science \cite{oh2025ultra, arteaga2019mueller}.

While these techniques can achieve exceptional precision under optimal conditions \cite{arteaga2012mueller}, they are limited when acquiring together data with strong intensity variations. In the case of MM imaging, complex samples frequently can present a combination of dark areas and intense specular reflections  \cite{jiao2025analysis, li2025underwater}. In MM spectroscopy, dynamic range issues often come from the emission profile of the light source itself, as standard lamps frequently exhibit sharp, high-intensity spectral peaks alongside regions of very low emission across the electromagnetic spectrum. Dynamic range limitations arise because, in most modern systems designed for fast acquisition, exposure time is not adjusted pixel by pixel or for each spectral point. Instead, the instrument is typically configured with a single acquisition setting for the entire measurement, which limits its ability to capture both very weak and very strong signals simultaneously.

These limitations are directly tied to the inherent dynamic range of the digital sensors employed (such as CCD based cameras or spectrometers). The dynamic range is bounded by the sensor's full well capacity at the high end and its read noise floor at the low end \cite{holst2007cmos}. If exposure times are increased to capture sufficient signal in dark or highly depolarizing regions, the sensor pixels saturate in intensely bright areas (such as specular reflections or lamp emission peaks). This saturation causes severe artifacts in the calculated MM. Alternatively, if the exposure is reduced to avoid saturation, the signal in dark regions falls below the noise floor, degrading the signal-to-noise ratio (SNR).

In imaging, various strategies have been proposed to avoid these intensity discrepancies. Some methodologies involve physically reorienting the sample to deflect specular reflections away from the detector \cite{xie2025reciprocal}; however, this restricts the permissible measurement geometries and is not always physically viable. Other approaches, such as implementing dual-camera systems with beam splitters \cite{evtushenko2012high} or using non-optimal conditions \cite{pardo2025method}, significantly increase the cost, alignment complexity, or spatial data loss. In spectroscopy, the choice of the illumination source is typically considered critical. Incoherent light sources with relatively smooth emission spectra, without pronounced spectral peaks within the wavelength region of interest, are generally preferred to minimize dynamic range issues.

The concept of high dynamic range (HDR) has become very popular in recent years, for example in digital photography and particularly in mobile phone imaging. Scenes with large intensity variations are often captured using HDR algorithms, which combine multiple exposures to extend the effective dynamic range of the recorded image \cite{banterle2017advanced}. While highly effective for visualization, these computational algorithms rely on a nonlinear mathematical process to compress the dynamic range for visual display \cite{debevec2023recovering, reinhard2023photographic}. Because MM extraction relies on the linearity of the measured intensity states, introducing a nonlinear algorithm generally invalidates the polarimetric demodulation process. 

To address this challenge without compromising the integrity of the polarimetric measurements, this article presents a straightforward HDR methodology applicable to multiple polarimetry setups that does not require complex nonlinear algorithms. The proposed technique relies on the direct addition of raw intensities captured at multiple exposure times prior to the calculation of the MM. Because each spatial pixel (in MM imaging) or spectral wavelength (in MM ellipsometry) is processed independently, we can selectively accumulate intensity data based on the specific response of that element. This strategy is particularly advantageous in systems where the integration time is decoupled from the polarization modulation, such as step-scan configurations with rotating compensators, allowing exposure settings to be freely adjusted without constraints imposed by the polarization-modulation hardware.

By extending the detector's effective well depth, this technique enables calculation of the 16 MM elements with significantly improved SNR in low-light regions while eliminating artifacts caused by saturation. Consequently, this approach effectively reduces the severe measurement errors caused by the sensors' restricted dynamic range. While other inherent limitations of the detector, such as minor deviations in sensor linearity or pixel cross-talk, will naturally persist alongside standard optically induced errors, this technique removes the primary electronic limitation of full-well capacity. This ensures the instrument can capture high-fidelity polarimetric data across extreme intensity gradients without compromising the accuracy of the extracted MM.

To demonstrate the versatility and robustness of this HDR technique, we integrated it into two distinct hardware configurations. The first is a broadband spectroscopic MM ellipsometer based on discrete-angle rotating Fresnel rhomb compensators \cite{bian2021mueller,bian2022calibration}, which measures across a highly variable lamp emission spectrum. The second is a widefield MM imaging polarimeter operating in a backscattering geometry \cite{pardo2024wide}, used to measure samples with highly varying intensities. Under these conditions, in both modalities, the proposed technique successfully eliminates saturation artifacts while preserving good quality data in low-signal regions.
\section{Theory description}

\subsection{Mueller matrix polarimetry}

MM polarimetry is described by Stokes-Mueller calculus. For any type of polarimetry architecture, the polarization state of light is represented by a $4 \times 1$ Stokes vector ($\mathbf{S}$), and the properties of the sample are described by a $4\times4$ Mueller matrix ($\mathbf{M}$), which acts as the transformation between the incident light ($\mathbf{{S}_{in}}$) and the output light ($\mathbf{{S }_{out}}$) \cite{chipman2018polarized}.

\begin{equation}
   {\mathbf{{S }_{out}}}=\mathbf{M} {\mathbf{{S}_{in}}}
\end{equation}

In a typical system, a Polarization State Generator (PSG) creates a sequence of known input polarization states, and a Polarization State Analyzer (PSA) projects the light interacting with the sample onto a sequence of known analysis states.  For any given configuration of the PSG and PSA, the optical detector of a common system measures a raw scalar intensity, $I_k$. To extract all 16 elements of the sample's MM, the system must capture a set of $N$ discrete intensity measurements ($N \geq 16$) using different combinations of generation and analysis states. Mathematically, this is expressed as a system of linear equations. By reshaping the $4 \times 4$ matrix $\mathbf{M}$ into a $16 \times 1$ vector $\vec{\mathbf{M}}$, and arranging the $N$ measured intensities into an $N \times 1$ vector $\mathbf{I}$, the polarimetric measurement process is defined by:

\begin{equation}
   {\mathbf{I}}=\mathbf{W}{\vec{\mathbf{M}}}
\end{equation}

Here, $\mathbf{W}$ is the $N \times 16$ basis matrix, defined entirely by the calibration of the PSG and PSA states. To retrieve the Mueller matrix, this linear system must be inverted, typically using the pseudoinverse $\mathbf{W}^{+}$.

\begin{equation}
    \mathbf{M} = \mathbf{W}^{+} \mathbf{I}
\end{equation}

This equation highlights a critical constraint in polarimetry: the extraction of the Mueller matrix relies on a strictly linear relationship between the physical Mueller matrix $\mathbf{M}$ and the measured intensity vector $\mathbf{I}$ \cite{mazumder2023optical}.

\subsection{Dynamic range limitations of the sensor}

The validity of the linear inversion depends entirely on the accuracy of the vector $\mathbf{I}$. In a physical detector (such as a CCD/CMOS camera or spectrometer), the measured digital intensity $I_k$ is the result of integrating the incident light over a specific exposure time $t$. If the amount of light that arrives at the detector is low, the resulting $I_k$ is dominated by the detector's read noise, yielding a poor SNR. If it is too large, the number of accumulated photons exceeds the pixel's full-well capacity during the exposure time $t$. Consequently, the measured signal saturates, and it is artificially capped at the sensor's maximum digital limit.

To understand this limitation, it is necessary to consider the physical architecture of the digital optical sensors (typically CCD or CMOS cameras or spectrometer arrays) used in polarimetric measurements. Each active element or pixel in a sensor acts as a microscopic well that collects photoelectrons generated by incident photons during the exposure time $t$.

The maximum charge a pixel can hold is known as its full well capacity. Once this physical limit is reached, the pixel cannot detect any additional light. Consequently, the sensor outputs its maximum digital value (for example, 255 for an 8-bit sensor or 4095 for a 12-bit sensor), resulting in a saturated, clipped measurement that no longer represents the true light intensity. On the other hand, the minimum detectable light signal is dictated by the sensor's noise floor, which primarily consists of read noise and dark current. If the incident light intensity is extremely low, the generated photoelectrons become indistinguishable from electronic noise, yielding a critically low SNR.

The intrinsic dynamic range of a detector is defined as the ratio between its full well capacity and its noise floor. When measuring a polarimetric scene where the contrast between the brightest area (such as a specular reflection or a spectral lamp peak) and the darkest region exceeds the sensor's dynamic range, a single exposure time $t$ cannot accurately capture both extremes. Adjusting $t$ forces a choice between sacrificing the dark regions to the noise floor or corrupting the bright regions through saturation.

\subsection{HDR enhancement of polarimetry measurements through addition of intensities}

In standard MM polarimetry, the conventional procedure is to acquire a set of $N$ intensity measurements corresponding to $N$ polarization states generated by the PSG and analyzed by the PSA. These intensity measurements are captured using a single, carefully selected exposure time that maximizes data quality across the largest possible area of the sample. The goal is to ensure the regions of interest use the maximum dynamic range without saturating. However, when a whole image or spectrum is acquired simultaneously, it is often unavoidable that some pixels or spectral regions remain underexposed and limited by dynamic range.

We propose a new multi-exposure approach in which the $N$ polarization states are measured across $P$ different exposure times. By directly summing the raw intensities from multiple exposures, we can artificially extend the detector's full well capacity. To maintain the linear proportionality required for the MM inversion, the effective exposure times for a given pixel must remain identical across all $N$ measurements for the different polarization states. Therefore, if a specific element (a spatial pixel or a wavelength channel) saturates at a given exposure time in any of the $N$ polarization states, that specific exposure time is discarded for that element across all polarization states, even if it did not saturate in the others. This masking approach can be presented as,
\begin{equation}
    w_p = \prod_{n=1}^{N} \mathcal{B}(I_{n,p})
\end{equation}
where $\mathcal{B}(I)$ is a boolean function defined as:
\begin{equation}
    \mathcal{B}(I) = 
    \begin{cases} 
    1 & \text{if } I < I_{sat} \\ 
    0 & \text{if } I \geq I_{sat} 
    \end{cases}
\end{equation}

Once the invalid exposure times are masked out for each pixel, the final HDR intensity image (or spectrum) for the $n$-th polarization state is calculated as the sum of the valid intensities:
\begin{equation}
    I_{n}^{HDR}=\sum_{p=0}^P w_p I_{n,p}
    \label{eq:HDR}
\end{equation}
where $n$ represents the specific polarization state (from $1$ to $N$), $p$ is the index of the evaluated exposure time (from $1$ to $P$), $I_{n,p}$ is the raw intensity measured for that specific state and exposure time and $w_p$ is a binary mask for each pixel of  $I_{n,p}$: $w_p = 0$ if exposure $p$ caused saturation in any of the $N$ states, and $w_p = 1$ if the exposure remained completely unsaturated. By utilizing the same mask $w_p$ for all $N$ states, the effective integration time remains constant across the measurement set, preserving the necessary linear relationship for polarimetric decomposition. 

To successfully implement this technique, the user must  select a sequence of exposure times that ensures sufficiently high-quality intensity data across the extremes of the sample's response. It is important to note that, due to the nature of the reconstruction, a correct representation of the $M_{00}$ element of the MM is not possible, as each area of the measurement will use different exposure times depending on its intensity. Nevertheless, an accurate determination of all the elements of the normalized MM will be obtained.

\section{Experimental setups}

To validate the practical implementation of the proposed HDR acquisition technique, we applied it to two distinct pre-existing setups in our laboratory: a spectroscopic Mueller matrix (MM) ellipsometer and a macroscopic MM imaging polarimeter. In both systems, the exposure time is decoupled from the polarization modulation process, as they are based on dual-rotating compensators operating in a step-scan configuration, so that no optical component is moving during data acquisition.

The first setup is a broadband MM ellipsometer based on discrete-angle rotating Fresnel rhomb compensators, as detailed by Bian et al. \cite{bian2021mueller}. The use of Fresnel rhombs provides nearly achromatic retardation across a broad spectral range. However, some light sources used in such spectroscopic systems typically exhibit highly uneven emission spectra, characterized by intense emission peaks alongside regions of very low intensity. This spectral disparity makes it an ideal candidate for our HDR technique, as long exposure times are required to achieve a viable SNR in the low-emission spectral regions, which inherently saturate the detector at the high-intensity peaks.

The second configuration is a macroscopic MM imaging polarimeter operating in a backscattering geometry, using a discrete-angle rotation architecture described by Pardo et al. \cite{pardo2024wide}. In backscattering imaging, samples with complex topographies or moist surfaces produce intense specular reflections that saturate the camera sensor, corrupting the polarimetric data in those localized regions. Applying HDR to this setup preserves high-fidelity data in low-intensity areas of the sample without losing the data beneath the specular highlights.

To process data from both configurations, we implemented the linear addition HDR algorithm. In both setups, at least two exposure times were used to obtain intensity images for each polarization state.

\begin{figure}[H]
\makebox[\textwidth][c]{\includegraphics[width=16cm]{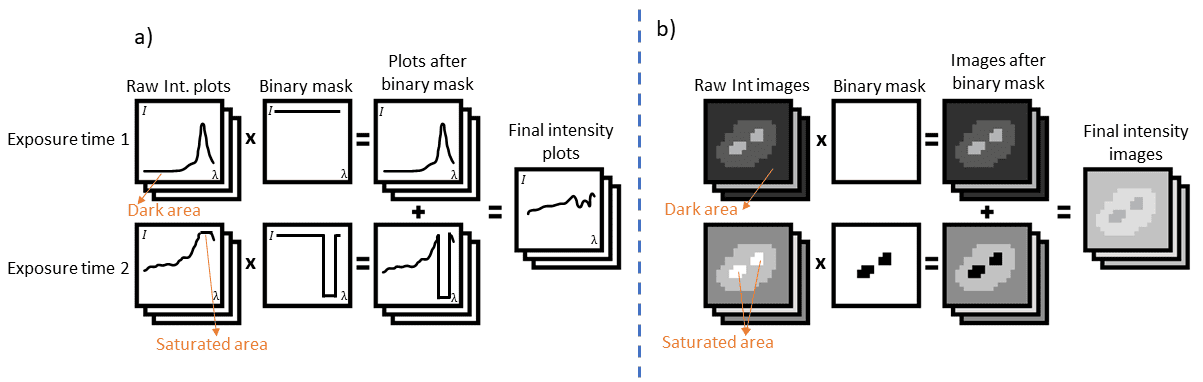}}
\caption{Flowchart of the linear HDR  methodology from Eq. \ref{eq:HDR} for (a) spectroscopic ellipsometry and (b) imaging polarimetry. Raw intensity measurements acquired at $P$ different exposure times are multiplied by binary masks to eliminate saturated elements. The valid data is then linearly summed to produce the final HDR intensities for each of the $N$ polarization states.  }
\label{fig:hdr_scheme}
\end{figure}

Fig. \ref{fig:hdr_scheme} displays the flowchart of our HDR data processing workflow. The process begins by capturing raw intensity data at $P$ distinct exposure times for each of the $N$ polarization states. As shown in the chart, a computational binary mask is dynamically generated to identify saturation. If an element (a spatial pixel or a wavelength channel) reaches the sensor's full well capacity at a specific exposure time in any of the $N$ states, that exposure time is masked to zero for that specific element across all polarization states to preserve the linearity across the $N$ polarization states. Finally, the valid, unsaturated intensities are linearly summed and normalized by their combined valid exposure times. This yields the final effective intensity vector, $\mathbf{I_{HDR}}$, which is subsequently inverted to retrieve the MM.

\section{Results and discussion}
\subsection{Mueller matrix ellipsometer}

\begin{figure}[H]
    \centering
    \includegraphics[width=1\linewidth]{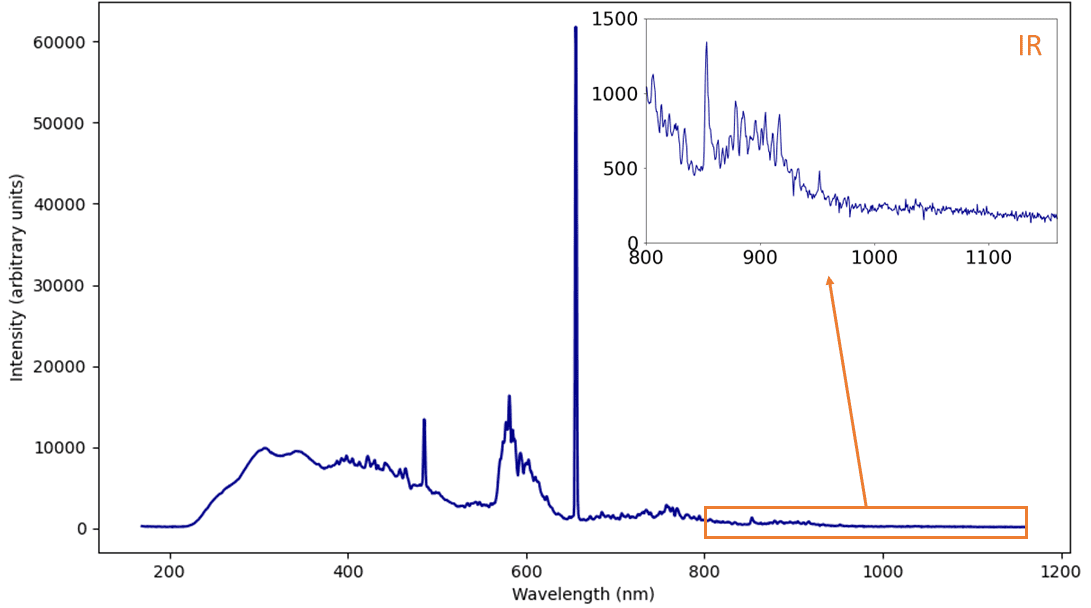}
    \caption{Deuterium lamp emission spectrum spanning the near UV, VIS, and NIR regions. The spectrum exhibits pronounced high-intensity peaks in the VIS and a significantly reduced emission level in the NIR, illustrating the dynamic range challenge addressed by the HDR approach. The Mueller matrix of the sample was evaluated over this entire spectral range. The inset (top right) shows a zoom of the NIR region, which is the spectral interval analyzed in detail in Fig.~\ref{fig:spahire800}.}
    \label{fig:deutspec}
\end{figure}

Using the MM ellipsometer, we measured the Mueller matrix (MM) of an $x$-cut sapphire crystal (1 mm thickness, double-sided polished) in reflection at an angle of incidence of 65°. The crystal is optically homogeneous across its surface; however, in this experiment, we deliberately employed a deuterium lamp as the illumination source, which exhibits strong spectral non-uniformity over the ultraviolet, visible, and near-infrared (UV–VIS–NIR) range (Fig.~\ref{fig:deutspec}). In particular, the emission becomes very weak in the NIR, while in the VIS the deuterium lamp presents sharp spectral peaks. Owing to this combination of low overall intensity in the VIS and NIR regions and pronounced emission lines, deuterium lamps are typically not used for spectroscopy beyond the UV. Nevertheless, our HDR approach overcomes these limitations, enabling reliable measurements across the full spectral range.

\begin{figure}[H]
    \makebox[\textwidth][c]{\includegraphics[width=15cm]{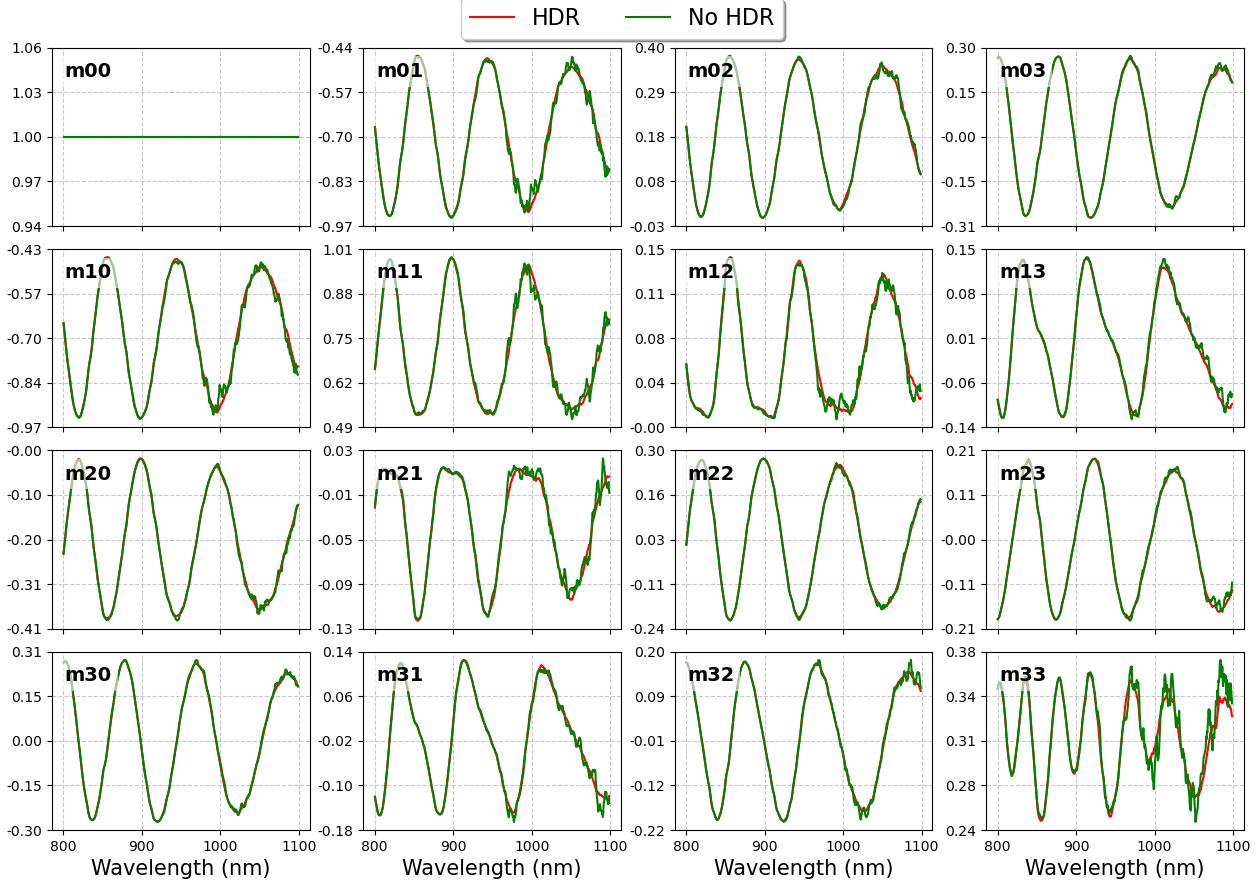}}
    \caption{Zoom of the NIR region in the normalized MM of the sapphire crystal measured with a deuterium lamp. The standard single-exposure result (green) exhibits significant noise due to the low source intensity, whereas the HDR measurement (red) shows a clear improvement in SNR. The high-frequency oscillations in the Mueller matrix arise from the wavelength-dependent birefringent phase accumulated over the 1 mm thickness of the double-sided polished crystal.}
    \label{fig:spahire800}
\end{figure}

To address this limitation, we applied our linear HDR approach using two distinct exposure times. As shown in Fig.~\ref{fig:spahire800}, this strategy leads to a substantial improvement in the SNR in the NIR region. In contrast, within the visible range, the HDR results closely match those obtained from a standard single-exposure measurement. This is expected, as the exposure time in the conventional approach is chosen to avoid detector saturation at the lamp’s intense emission peaks. While this constraint inevitably degrades data quality in the low-intensity NIR region, the multi-exposure HDR method enables sufficient signal accumulation in these spectral regions without compromising the integrity of the oscillatory behavior in the MM elements arising from the bulk birefringence of the crystal.

\subsection{Mueller matrix imaging polarimeter}

Using the imaging polarimeter, we evaluated two different scenes. 

\begin{figure}[H]

    \makebox[\textwidth][c]{\includegraphics[width=12.5 cm]{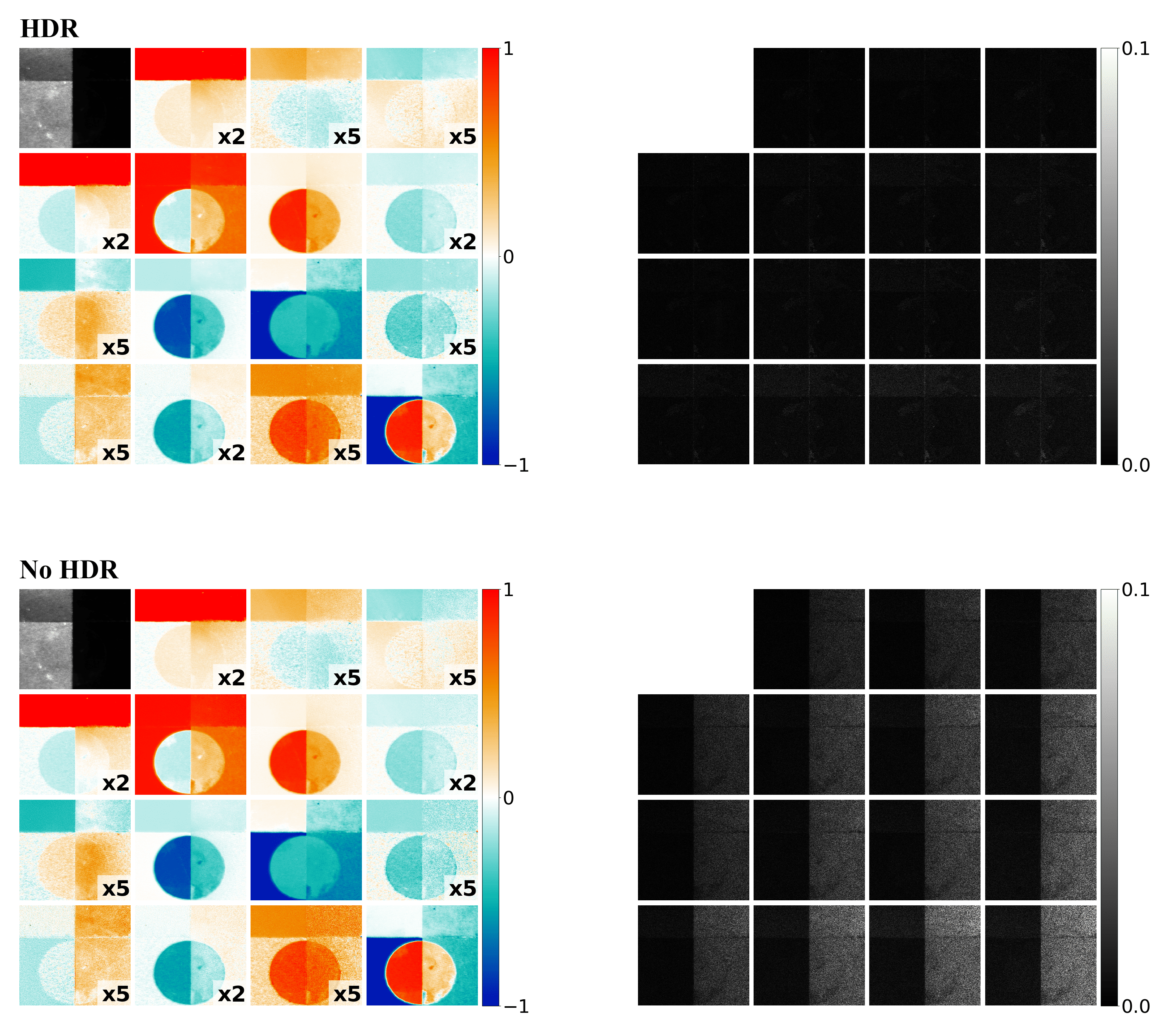}}
    \caption{Normalized MMs with off-diagonal scaled (left column) and SD of said MMs (right column) for both the HDR and no-HDR measurements (up and down, respectively) of our scene. In the top left corner of the MM, we can see the unnormalized $M_{00}$ calculated with a single exposure time. The SDs were calculated by doing 3 consecutive measurements. }
    \label{fig:sceneMM}
\end{figure}

Fig. \ref{fig:sceneMM} shows a scene composed of standard polarization elements—a rectangular film polarizer on the top and a circular film retarder on the bottom—placed over two distinct backgrounds: a highly reflective metallic surface on the left and a dark fabric on the right. These two sides of the scene present very different intensity responses. When the exposure time is adjusted to avoid saturation on the metallic surface, the signal from the dark region becomes severely underexposed, leading to a poor SNR. As in the spectroscopic case, applying our HDR technique using multiple exposure times effectively reduces noise in these low-signal regions while avoiding saturation artifacts. This improvement is clearly demonstrated in the standard deviation (SD) image comparison in Fig. \ref{fig:sceneMM}.

\begin{figure}[H]

    \makebox[\textwidth][c]{\includegraphics[width=12.5 cm]{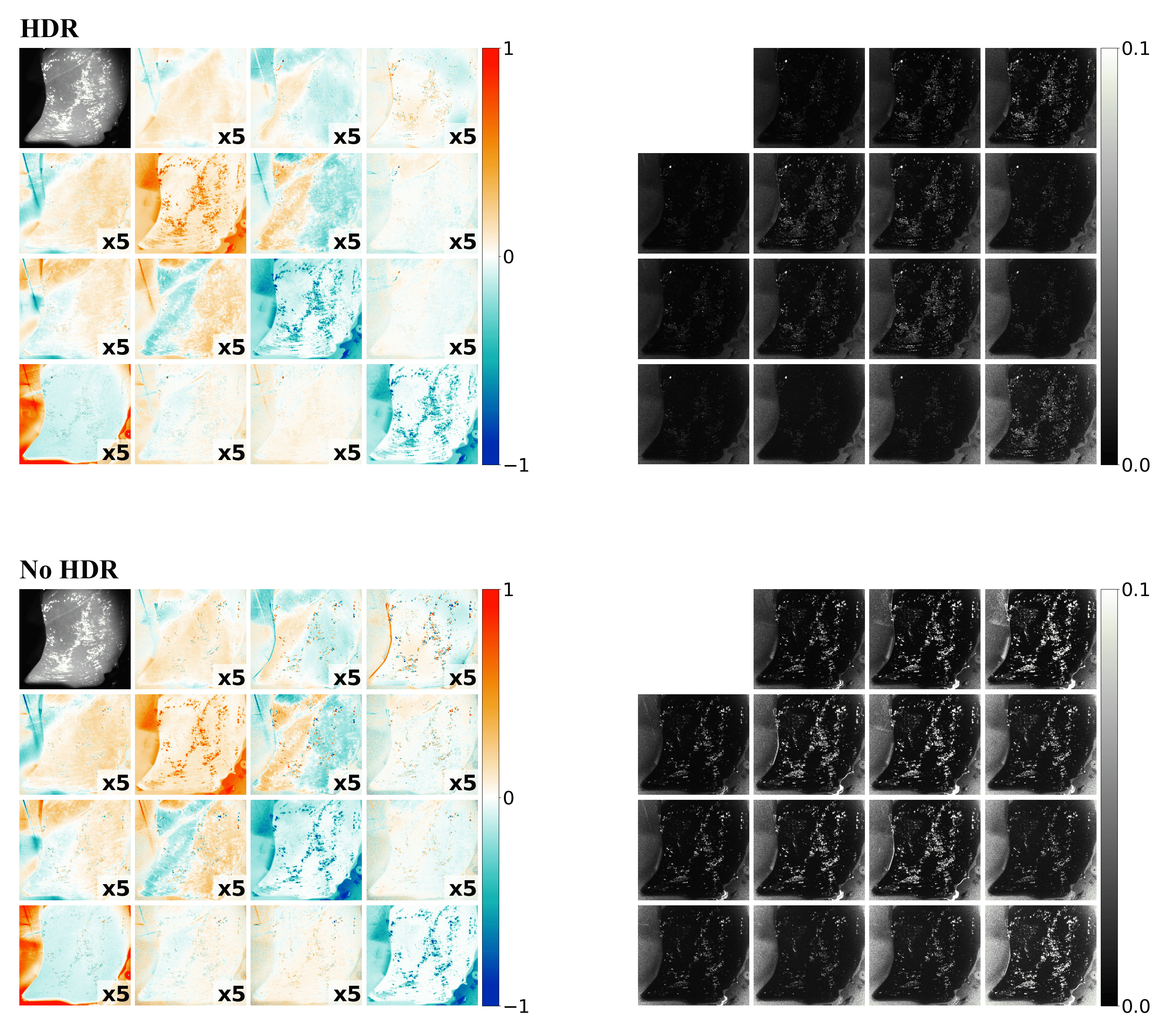}}
    \caption {Normalized MMs with off-diagonal scaled (left column) and SD of said MMs (right column) for both the HDR and no-HDR measurements of a chicken breast's biological sample (up and down, respectively). In the top left corner of the MM, we can see the unnormalized $M_{00}$ calculated with a single exposure time. The SDs were calculated by doing 3 consecutive measurements.}
    \label{fig:pollo5}
\end{figure}

To further demonstrate the versatility of our method, the second scene shows the MM of a chicken breast obtained from a local supermarket (Fig. \ref{fig:pollo5}). This example highlights a common challenge in biomedical polarimetry: biological tissues often exhibit surface-induced specular reflections  \cite{jiao2025probing}. When using an exposure time long enough to capture sufficient signal from the bulk of the tissue, these reflections inevitably cause the sensor to saturate, losing polarimetric data from these regions. By applying our HDR technique, we effectively reduce these saturation artifacts, allowing us to obtain more reliable polarimetric information across the entire surface. To quantify this improvement, we also calculated the SD of the MM. Comparing the HDR results (top) with the standard single-exposure data (bottom), the SD values are substantially lower for HDR, confirming a significant noise reduction.

%

%

%

\section{Conclusion}
In this work, we demonstrated the effectiveness of a linear HDR methodology to overcome the inherent dynamic range limitations of optical detectors in MM polarimetry. By capturing and accumulating data across multiple exposure times, the proposed technique successfully mitigates saturation-induced artifacts while substantially improving the SNR in low-intensity regions.

We validated this approach experimentally across two complementary modalities. In macroscopic imaging, the method effectively handles large intensity variations, suppressing specular reflections from biological tissues while enhancing low-signal regions in scenes with strong contrast. In spectroscopic ellipsometry, it enables accurate recovery of weak signals in low-emission spectral regions without inducing saturation at the intense emission peaks of the illumination source, potentially extending the usable spectral range of the instrument. Because the exposure time of CCD/CMOS detectors can often be rapidly adjusted, the method can be implemented without any hardware modifications and with negligible impact on acquisition time beyond the selected exposure settings.

By ensuring optimal exposure conditions for every spatial pixel or spectral wavelength, this technique guarantees the extraction of physically accurate polarimetric data. Because the method avoids nonlinear mathematical processing, it strictly preserves the proportionality required for the linear Stokes-Mueller inversion. This prevents the generation of artifacts and ensures the integrity of the calculated MM elements. Consequently, this approach offers a simple yet efficient solution for characterizing complex samples, eliminating the need for hardware modifications or software trade-offs. Future work will investigate using automated algorithms to dynamically determine the optimal sequence of exposure times, further simplifying the acquisition process for different samples.

\section*{Funding}
This work was supported by project PID2022-138699OB-I00 from the Ministerio de Ciencia e Innovación of Spain and by project 2024 PROD 00116 by Agència de Gestió d'Ajuts Universitaris i de Recerca (AGAUR).


\bibliographystyle{elsarticle-num-names} 
\bibliography{report}






\end{document}